\documentstyle[preprint,aps]{revtex}

\begin{document}
\tighten
\draft


\title{Intermediate Scales in Supersymmetric GUTs:\\ the Survival
 of the Fittest}

\author{Charanjit S. Aulakh$^{(1)}$, Borut Bajc$^{(2)}$,
Alejandra Melfo$^{(3)}$, Andrija Ra\v{s}in$^{(4,5)}$\\ and 
Goran Senjanovi\'c$^{(5)}$}

\address{$^{(1)}$ {\it Dept. of Physics, Panjab University, 
Chandigarh, India   }}
\address{$^{(2)}$ {\it J. Stefan Institute, 1001 Ljubljana, Slovenia}}
\address{$^{(3)}$ {\it CAT, Facultad de Ciencias, Universidad de
Los Andes, M\'erida, Venezuela }}
\address{$^{(4)}${\it Dept. of Physics and Astronomy, University of 
North Carolina, Chapel Hill, NC 27599, U.S.A. }}
\address{$^{(5)}${\it International Center for Theoretical Physics,
Trieste, Italy }}

\maketitle

\begin{abstract}
We show that intermediate scales in supersymmetric grand unified 
theories may exist naturally. Their origin is traced to the 
violation of the survival principle: in supersymmetry internal 
symmetries often forbid cubic couplings in the superpotential. 
This leads to a plethora of light supermultiplets whose 
masses are generated only by higher dimensional operators 
and thus suppressed by the cut-off scale. 
These new states often carry exotic quantum numbers and may be,
in some cases, accessible to experiments in the near future.

\end{abstract} 

\section{Introduction} 

Two main motivations for the low energy supersymmetry are the 
alleviation of the hierarchy problem and the successful unification 
of gauge couplings in the context of the minimal supersymmetric 
standard model (MSSM) \cite{mssm}. The quest for supersymmetry
has become one of the central issues of present day particle 
physics.

Today, on the other hand, we have growing observational evidence 
in  favor of non-vanishing neutrino masses. The most natural and 
the simplest way to generate  small neutrino masses is provided by 
the see-saw mechanism \cite{seesaw}. The accumulation of solar 
and atmospheric neutrino data points to the see-saw (right-handed 
neutrino) mass scale $M_R$ in the range of $10^{10}-10^{15}$ GeV 
\cite{smirnov99}. For believers in supersymmetry there is thus an 
important question as to whether there can be an
 intermediate scale 
 below the grand unification scale $M_X$. The
 conventional wisdom in 
the context of minimal grand unified theories
 without extra fine-tuning seems to be no. 

Recently, we have pointed out \cite{amrs99} an important connection 
between the renormalizable see-saw mechanism and low energy 
supersymmetry. 
First, if B$-$L symmetry is spontaneously broken, we can show that 
$R$-parity is an exact symmetry and the low-energy effective theory 
is the MSSM. Second, in many cases supersymmetry  is consistent 
with the canonical form of the see-saw, an important achievement. 
It is then essential to know whether we can have $M_R$ in the 
intermediate regime, as may be needed by neutrino data. In this 
letter we wish to show that, contrary to the conventional 
wisdom, it may be quite natural 
to have $M_R$ as an intermediate scale. In the minimal supersymmetric 
SO(10) theory $M_R$ can be as low as $10^{13}$ GeV with the prediction 
of lowering $M_X$ close to $10^{15}$ GeV, giving thus the old 
$d=6$ proton decay mode $p\to\pi^0e^+$ as the possibly dominant one. 
The reason for a different prediction from the minimal supersymmetric 
SU(5) stems from the violation of the survival principle \cite{survival} 
and it has generic features beyond any particular model. In other words, 
supersymmetric GUTs beyond 
SU(5) contain many potentially light supermultiplets which may 
modify the usual gauge coupling running. This is an important, 
albeit simple point, often overlooked in the literature (but not 
always, see for example \cite{light,cceel97}). Since it is generic to 
supersymmetric  gauge theories, we address it more carefully and illustrate 
it in a number of examples, starting from a minimal U(1) model, all the 
way to the supersymmetric  SO(10). In all cases we will find 
a violation of the naive survival principle. 

\section{Survival of the fittest} 

The survival principle is the natural assumption that a particle in 
multi scale theories takes the largest possible mass consistent with 
the symmetries of the theory. In particular, this means that the 
singlets under a particular gauge symmetry have masses that 
correspond to the scale of the larger symmetry under which they are 
not singlets. Of course, if they do not participate in symmetry breaking
and have a gauge invariant mass, then they can go to an
even higher scale. 

An important aspect of this principle is the usual manifestation of the 
Higgs mechanism in ordinary gauge theories. When a multiplet $\Phi$ 
develops a nonvanishing vacuum expectation value (VEV)
 through the potential 

\begin{equation}
V=-m^2\Phi^\dagger\Phi+\lambda^2(\Phi^\dagger\Phi)^2+...\;,
\label{vphi}
\end{equation}

\noindent
the Higgs scalars get masses of the order 

\begin{equation}
m_\Phi\approx\lambda\langle \Phi\rangle
\end{equation}

\noindent
except for the unphysical Higgs and the possible pseudo-Goldstone 
bosons in the case of accidental symmetries of the potential. Of 
course, in general we have more self couplings in (\ref{vphi}); 
$\lambda$ stands generically for all of them. Notice that there is no 
symmetry that forbids the coupling $\lambda$, in  as much as there is no 
symmetry that forbids the mass term. 
This is intimately related to the problem of hierarchies and thus we 
should not be surprised if the situation gets drastically different in 
supersymmetric  theories. 

In other words, unless one resorts to the unnatural fine-tuning  
$\lambda\ll 1$, we expect $m_\Phi\sim  \langle \Phi\rangle$. This 
looks rather natural, so natural, that it is normally taken for 
granted even in supersymmetry. It is however potentially wrong as 
we show now. In supersymmetry the $\Phi^4$ term in (\ref{vphi}) 
arises from the superpotential $W=\lambda\Phi^3/3+...$. However, 
the cubic term is often forbidden for symmetry reasons and so 
is the quartic self-coupling in the potential. In general there 
will be higher dimensional terms of the form 

\begin{equation}
W_{nr}={\Phi^{n+3}\over M^n}\;,
\end{equation}

\noindent
where $M$ is the cutoff scale of the theory; $n>0$ and it depends on 
the symmetry in question. Thus the $\lambda$ coupling is strongly 
suppressed, $\lambda_{\rm eff}\sim \langle \Phi\rangle^n/M^n$, and 

\begin{equation}
 m_\Phi\approx \lambda_{\rm eff}\langle \Phi\rangle 
 \approx{\langle\Phi\rangle^{n+1}\over M^n}\;.
\end{equation}

Even for $n=1$, we can have $m_\Phi\sim  \langle 
\Phi\rangle^2/M\ll  \langle \Phi \rangle$, 
since $\langle \Phi\rangle\ll M$. Here $\Phi$ stands for the whole 
super-multiplet of bosons and fermions, since we will  deal with 
$\langle \Phi\rangle \gg M_W$ when supersymmetry is assumed to be a good 
symmetry. This fact is clearly important for phenomenology, but 
also for the unification of couplings. If $M\sim  M_X$, and 
$\langle \Phi\rangle=M_I$, the intermediate scale, many particles 
will start contributing to running at much lower energies 
$M_I^2/M_X$. In what 
follows we give some examples of the above with  growing 
complexity and physical relevance. 

\subsection{A U(1) prototype example}

Take the simplest supersymmetric  anomaly-free U(1) gauge
 symmetry with two chiral superfields $\Phi$ and 
$\overline\Phi$ (charges $\pm 1$). The U(1) symmetry forbids cubic 
couplings and the renormalizable superpotential is trivial, 
$W = m\Phi\overline\Phi$. It implies no breaking of the symmetry,
$\langle \Phi\rangle = \langle\overline\Phi\rangle = 0$.
 In order  to have symmetry breaking, one needs at least a $d=4$ term 
in the superpotential,

\begin{equation}
W = m\Phi\overline\Phi+{(\Phi\overline\Phi)^2\over 2M}\;,
\end{equation}

\noindent
which together with 

\begin{equation}
V_D = {g^2\over 2}(|\Phi|^2-|\overline\Phi|^2)^2
\end{equation}

\noindent
provides nonvanishing VEVs: $\langle \Phi\rangle=\langle 
\overline\Phi\rangle=\sqrt{mM}$. 
Obviously, the effective coupling is suppressed 

\begin{equation}
\lambda\sim \langle \Phi\rangle/M\;.
\end{equation}

Whereas the combination $\Phi-\overline\Phi$ belongs to the super-Higgs 
multiplet (mass of order $g\langle \Phi\rangle$), $\Phi+\overline\Phi$ 
has a smaller mass 

\begin{equation}
m_{(\Phi+\bar\Phi)} \approx \langle \Phi\rangle^2/M\;. 
\end{equation}

This simple model portrays well the general 
situation we will find in the following.

\subsection{The supersymmetric Left-Right model}

The minimal such model mimics completely the above U(1) example. 
The gauge group is SU(2)$_L\times$ SU(2)$_R\times$ U(1)$_{B-L}$  
\cite{leftright}. We have left handed ($\Delta$, $\overline\Delta$) 
and right handed ($\Delta_c$, $\overline\Delta_c$) triplets and  in
the absence of nonrenormalizable terms there is no interaction 
in the superpotential \cite{amrs98}. Again, one needs nonrenormalizable 
terms to generate symmetry breaking. Just as above, except for 
the super-Higgs multiplet, the rest of the particles should have 
masses of the order $M_R^2/M$, where 
$M_R=\langle \Delta_c\rangle=\langle \overline\Delta_c\rangle$ 
is the symmetry breaking scale of parity and SU(2)$_R$ and $M$ 
is the cutoff. 

The neutral multiplets  in $\Delta_c$ and $\overline\Delta_c$ 
get a VEV, and the corresponding masses through the super-Higgs mechanism.
The states that do not belong to the super-Higgs multiplet are one
neutral complex combination and two doubly charged chiral multiplets
from $\Delta_c $ and $\overline\Delta_c$, and of course all the states
from $\Delta$ and $\overline\Delta$. 
This leads 
to an important prediction of light doubly charged supermultiplets 
potentially observable by experiment \cite{ams97}, with striking
signatures at future colliders\cite{dm98}. This result 
can be extended to include an arbitrary number of gauge singlet 
fields \cite{cm97}. In this case the lightness of the doubly charged 
multiplets can be traced to their pseudo-Goldstone nature \cite{cm97}. 
Namely, with  parity odd singlet(s) one can generate the VEVs for the 
right handed fields, but the renormalizable superpotential has a larger 
accidental symmetry and the doubly charged fields are the Goldstone 
modes. They get masses if one includes the higher dimensional 
terms which break the accidental symmetry. 

However, our point is much more general and is valid whenever one 
needs to invoke higher dimensional operators to generate relevant 
interactions. It says simply, as we already emphasized, that 
except for the super-Higgs mechanism, the rest of the supermultiplet 
masses are suppressed by the cut-off scale associated with 
the higher dimensional interactions. In the minimal model
(without singlets)  
the lightness of the doubly charged states has nothing to 
do with the pseudo-Goldstone mechanism: simply, in the absence 
of higher dimensional operators there is no symmetry breaking 
whatsoever. When they are included and L-R symmetry is broken, 
the doubly charged states are light because they do not 
belong to the super-Higgs multiplet.

\section{Supersymmetric SO(10) theory}

 From the point of view of unification this will be the simplest 
illustration of the above idea. The minimal renormalizable theory 
requires the symmetric  traceless ${\bf 54}$  supermultiplet 
 and the antisymmetric ${\bf 45}$  adjoint. This allows 
for the breaking of SO(10) down to SU(2)$_L\times $SU(2)$_R\times$ 
U(1)$_{B-L}\times$ SU(3)$_C$. Further breaking is achieved by either 
({\it i}) $ {\bf 126} $ and $ {\bf \overline{126}}$ or 
({\it ii}) {\bf 16 } and ${\bf  \overline{16}}$. 
The former turns out to be more interesting so we study 
it first in detail. 

\subsection{${\bf 126}$ and {${\bf \overline{126}}$} or 
renormalizable see-saw mechanism}

A renormalizable SO(10) theory with a see-saw requires the minimum 
set of Higgs representations which break SO(10) down to the MSSM 
\cite{leem94}

\begin{equation}
S = {\bf 54} \; , \quad A = {\bf 45} \; , \quad \Sigma = {\bf 126} \; , 
\quad \overline \Sigma = \overline{{\bf 126}} \;.
\label{higgses}
\end{equation}

Although SO(10) is anomaly-free, as is well-known, one has to use 
both $\Sigma$  and $\overline \Sigma$ in order to ensure the flatness 
of the D-piece of the potential at large scales $\gg M_W$. The most 
general superpotential containing $S, A, \Sigma $ and $\overline \Sigma$ 
is

\begin{eqnarray}
W &= &{m_S \over 2} {\rm Tr}\, S^2 + {\lambda_S \over 3} {\rm Tr}\, S^3 
+ {m_A \over 2} {\rm Tr}\, A^2  + \lambda {\rm Tr}\, A^2S \nonumber \\
& & + m_\Sigma \Sigma \overline\Sigma + \eta_S \Sigma^2 S +
 \overline\eta_S 
{\overline\Sigma}^2 S + \eta_A \Sigma\overline\Sigma A\;.
\label{superpot}
\end{eqnarray}

We wish to obtain the pattern of symmetry breaking

\begin{eqnarray}
SO(10) & \stackrel{\langle  S \rangle = M_X }{\longrightarrow} & 
SU(2)_L\times SU(2)_R\times SU(4)_C  \nonumber \\
& \stackrel{\langle  A \rangle = M_C }{\longrightarrow} &
SU(2)_L\times SU(2)_R\times U(1)_{B-L}\times SU(3)_C \nonumber \\
& \stackrel{\langle  \Sigma \rangle = \langle  {\overline 
\Sigma} \rangle = M_R }{\longrightarrow} & 
SU(2)_L\times U(1)_{Y}\times SU(3)_C
\label{sixteen} 
\end{eqnarray}

\noindent
and this is achieved with non-vanishing VEVs in the directions 

\begin{equation}
s = \langle  (1,1,1)_S \rangle   \quad 
a = \langle  (1,1,15)_A \rangle \quad 
b = \langle  (1,3,1)_A \rangle \quad 
\sigma = \langle  (1,3,10)_\Sigma \rangle \quad 
\overline\sigma = \langle  (1,3,\overline{10})_{\overline\Sigma}
\rangle \quad 
\label{vevnames}
\end{equation}

\noindent
(where here and in the following we label fields by their 
SU(2)$_L \times$ SU(2)$_R \times$ SU(4)$_C$ quantum numbers). 
Notice that $\langle S \rangle$ also preserves the discrete
parity symmetry $D_{LR}$ \cite{parida}, which is then broken in the
next step by the parity-odd singlet in $(1,1,15)_A$. This will
of course cause the left-right inequality of masses already at
the scale $M_C$.

The F-flatness equations are 

\begin{eqnarray}
m_S s + {1 \over 2} \lambda_S s^2 + {2\over 5} 
\lambda (a^2 -b^2) &=& 0 \;, \nonumber \\ 
a (m_A  + 2 \lambda s) + {1 \over 2} \eta_A \sigma\overline\sigma
= b (m_A  - 3 \lambda   s) + {1\over 2} \eta_A \sigma \overline\sigma 
&=& 0 \;, \nonumber \\
\sigma \left[ m_\Sigma + \eta_A (3 a + 2 b) \right]
&=& 0 
\label{efeqs}
\end{eqnarray}

\noindent 
and $\sigma = \overline\sigma $ guarantees D-flatness. 

In order to get the required multi-scale symmetry breaking one needs 
$s\gg a \gg \sigma= \overline\sigma \gg b \simeq \sigma^2/s$. 
This can be achieved by paying the usual price of fine-tuning 

\begin{equation}
m_A + 2 \lambda s \simeq {\sigma^2 \over a} \ll s\;,
\label{finea}
\end{equation}

\noindent
which then ensures

\begin{equation}
b \simeq {\sigma^2 \over s} \ll \sigma
\label{finea2}
\end{equation}

\noindent 
(it is important to keep in mind that $b$ can never vanish). 
A comment is in order. In the F flatness conditions (\ref{efeqs}) 
we ignore the fields in {\bf 16} and {\bf 10} 
dimensional representations. 
This  is justifiable for the Standard Model non-singlet fields, 
but not for $\tilde{\nu^c}$ in {\bf 16}. However, in the vacuum 
$\sigma = \overline \sigma$, $\langle \tilde{\nu^c}\rangle = 0$, 
as can be easily proved following the similar analysis in the 
context of LR theories \cite{amrs98}. 

We now compute the spectrum of the 
theory. This has been discussed
in  reference\cite{leem94}, however only in the limit of a 
single step breaking of SO(10). In this limit one misses the 
possibility of light states which have masses suppressed by the 
ratios of different scales of symmetry breaking. In what follows 
we identify all such states together with the rest of the 
spectrum\footnote{For previous attempts in supersymmetric
SO(10) theories with 126 fields see \cite{aulamoha,sato},
where in \cite{aulamoha} the presence of light states
was noticed. However, the effects of $M_I^2/M$ that are central
to our paper were missed.}.

At $M_X$, $(2,2,6)_S$ gets a mass of the order $\langle S\rangle$ 
due to the super-Higgs mechanism. So do all the remaining fields 
in $S$ and almost all of $A$, through the terms Tr$S^3$ and Tr$SA^2$. 
The sole exception 
are the $(1,1,15)$ fields in $A$, which we choose to have a much smaller 
mass imposing the fine-tuning condition (\ref{finea}): this multiplet 
will thus get a VEV in the next stage of symmetry breaking, at a much 
lower scale. Couplings with $S$ also give a 
mass $\sim M_X$ to all fields in $\Sigma$ and $\overline \Sigma$, 
with the exception of the SU(4) decuplets. There is no 
fine-tuning involved here; this is automatic, since $S$ is 
coupled only to $\Sigma^2$ or $\overline{\Sigma}^2$ and not to 
$\Sigma\overline{\Sigma}$. 

In the next stage $(1,1,15)_A$ will get a nonvanishing VEV in the 
direction of the SU(3) color singlet. Again, the super-Higgs 
mechanism gives a mass $M_C$ to the colored triplets in $(1,1,15)_A$. 
The couplings with $A$ provide masses at this scale to the left-handed 
components $(3,1,\overline{10})_\Sigma$ and $(3,1,10)_{\overline\Sigma}$, 
and to the colored fields in their right-handed analogues. As mentioned
before the reason for the asymmetry in the left-right masses already
at this scale comes from the fact that the singlet in $(1,1,15)_A$
is parity-odd. But as we 
anticipated, the color octet and the singlet in $A$ will survive the 
Pati-Salam breaking. The reason once again can be traced to the 
superpotential (\ref{superpot}).  Both $A$ and the $\Sigma$ 
($\overline\Sigma $) fields lack cubic self-interactions. 
In order to get effective interactions one has to integrate out the 
heavy field $S$. Notice that the color octets may get a 
contribution of the order $M_R^2/M_C$, which is just due to 
the necessary fine-tuning of formula (\ref{finea}). 

At this point the situation resembles the Left-Right model discussed 
above. At the scale $M_R$, the  SU(2)$_R$ triplets in $\Sigma$ and 
$\overline\Sigma$ get their VEVs, but two doubly-charged chiral 
multiplet and one neutral combination remain light. The mass spectrum 
is summarized in the Table below.

\begin{center}
\framebox{\begin{tabular}{l|c}
\hspace{2cm}State &  Mass \\
\hline 
\begin{tabular}{l}
all of $S$ in ${\bf 54}$ \\
all of $A$ in ${\bf 45}$, except $(1,1,15)_A$\\
all of $\Sigma$ in {\bf 126} + $\overline \Sigma$ 
in ${\bf \overline{126}}$, except  SU(4)$_C$ decuplets 
\end{tabular}
&  $\sim M_X$ \\
\hline
\begin{tabular}{l}
$(3,1,\overline{10})_\Sigma$  + $(3,1,{10})_{\overline\Sigma}$  \\
color triplets and sextets of $(1,3,{10})_\Sigma$  and 
 $(1,3,\overline{10})_{\overline\Sigma}$  \\
color triplets of $(1,1,15)_A$
\end{tabular}
&  $\sim M_C$ \\
\hline
\begin{tabular}{l} 
$(\delta_c^0 - \overline{\delta_c^0}), \delta_c^+,\overline{\delta_c^+}  $\\
\hspace{0.5cm} from the color singlets of $(1,3,{10})_\Sigma$
and $(1,3,\overline{10})_{\overline\Sigma}$ 
\end{tabular}
& $\sim M_{R}$ \\
\hline
color octet and singlet of $(1,1,15)_A$
& $\sim Max\left[ {M_R^2 /M_C}, {M_C^2 /M_X} \right]$ \\
\hline
\begin{tabular}{l}
$(\delta_c^0 + \overline{\delta_c^0}) \; , \;\delta_c^{++},\overline{
\delta_c^{++}}  $\\
\hspace{0.5cm} from the color singlets of $(1,3,{10})_\Sigma$ 
and $(1,3,\overline{10})_{\overline\Sigma}$ 
\end{tabular}
& $\sim {M_{R}^2 / M_X}$
\end{tabular}}
\vspace{0.5cm}
\end{center}

\vspace{0.5cm}

In addition, we have to include a {\bf 10}
 representation, which  contains the  MSSM Higgs doublets. Couplings with $S$ (and the usual
 fine-tuning) will ensure that the colored fields in {\bf 10} get
 a mass $\sim M_X$. 
   However, as is well known, at least 
two ten-dimensional Higgs representations are needed in general 
in order to generate non-vanishing quark mixing angles in the 
CKM matrix. This implies two extra Higgs doublets, which get 
a mass through the VEV $b$ in $A$ of the order $M_R^2/M_X$. 
The remaining fields in {\bf 10} and three generations of 
{\bf 16} (with the exception of
 the right-handed neutrinos) get their mass at $M_W$.

The Table clearly illustrates our point about the violation of the 
survival principle. The last two rows contain  a number of states 
with masses below the corresponding scales of symmetry breaking. 
The other entries of the Table are clear and conform to the 
survival hypothesis. If one is weary of effective field theory 
arguments, it is a straightforward exercise to obtain the masses 
by direct calculation. For example the color octet states 
in $(1,1,15)_A$ mix with their super heavy counterparts 
in $(1,1,20)_S$; hence their see-saw-like suppressed mass.

The presence of these new states significantly alters the 
unification predictions. We wish to have a rough qualitative 
estimate of the new mass scales and thus it is only appropriate 
to perform this at the one-loop level. It is straightforward to
write down the renormalization group equations and  
solve for $M_X$, $M_C$ and $M_R$ in terms of the (unknown) unified 
coupling $\alpha_U$. We find that the most interesting scenario 
occurs for $M_C^2/M_X > M_R^2/M_C$ (which fixes the mass of the 
octet). 

We find, remarkably enough, that lowering 
$M_R$ implies the lowering of $M_X$.
This is an important result which changes the nature 
of the proton decay.
  Clearly, the proton decay constraints set a lower limit 
on $M_X$.
At the same time we make sure  that the value of
 $\alpha_U^{-1}$ remains perturbative. 
With the scale of supersymmetry 
breaking $\sim M_Z$, requiring $M_X \gtrsim 10^{15.5}$ GeV gives 
\begin{equation}
M_C \gtrsim 10^{14.7}{\rm GeV} \;\; ,\quad M_R 
\gtrsim 10^{13.5}{\rm GeV} \;\; ,
\end{equation}
for $\alpha_3(M_Z)=0.119$. The lower limit on $M_R$ (and $M_C$)
gets further decreased for higher $\alpha_3(M_Z)$, and for
a typical MSSM value of $\alpha_3(M_Z)=0.126$ we get
$M_C \gtrsim 10^{14.3}$GeV, $M_R \gtrsim 10^{13.0}$GeV.

Notice that the scale $M_C^2/M_X$ is actually bigger 
than $M_R$. In the Table we have shown it below
only in order to separate new non-renormalizable 
mass scales from the original ones.
The precise value of an intermediate scale $M_R$ should be 
checked at the two-loop level including threshold effects 
(however see \cite{sher89}), since it is not too far 
from the unification scale. Its existence is clearly 
important for neutrino masses through the see-saw, 
but its impact is even more dramatic on the proton decay 
predictions. If $M_R$ lies below $10^{14}$ GeV, the dominant 
proton decay mode may as well be into the positron and the 
neutral pion, as in the non-supersymmetric theories. This 
is in sharp contradiction with the minimal supersymmetric 
SU(5) theory in which this mode is highly suppressed and 
dominant modes would necessarily involve kaons, not pions 
(for a complete analysis and references see \cite{mura93}). 
Of course, we may still have the kaon modes, however not 
necessarily dominant.

Anyway, the above example illustrates nicely our point: 
the combined effect of supersymmetry and internal 
symmetries may lead naturally to particles with masses 
much below the scale of symmetry breaking. Their impact 
on unification predictions may be non-trivial and should 
not be ignored. This is a dynamical question, though, 
that depends on the model in question. In the next 
case it will turn out, in spite of the presence of 
potentially light states, that the theory chooses 
the minimal SU(5) route without any intermediate scale.

\subsection{${\bf 16}$ and ${\bf \overline{16}}$ or large-scale 
R-parity breaking}

Supersymmetric  SO(10) models have been studied at length 
with the non-renormalizable version of the see-saw mechanism\cite{sedam}. 
More precisely, one chooses one (or more) 
pair of ``Higgs'' in the spinorial representation ${\bf 16} $
and $\overline{{\bf 16}}$ whose VEVs induce B-L breaking and the mass 
for the right-handed neutrino through the $d=4$ terms: 
$m_{\nu_R} \simeq \langle  {\bf 16} \overline {\bf 16}\rangle 
/M_{Pl}$. This enables one to get rid of the large 126-dimensional 
representations which render the high-energy behavior above 
$M_X$ non asymptotically free. 

The disadvantage of this program, though, is that then R-parity 
is broken at a large scale $M_R$, and thus one needs additional, 
ad-hoc symmetries to understand the smallness of R-parity breaking 
at low energies. This appears to us  more problematic than 
the loss of asymptotic freedom at super high energies.   
Even more important, it will turn out that, in spite of the possibly 
light particles, the unification constraints push all the scales 
towards $M_X$, leaving no room for intermediate scales. We still 
include this example in order to show the subtlety of the issue: 
it is a dynamical question whether or not one can have intermediate
mass scales. 

The superpotential closely resembles the previous case's one 

\begin{equation}
W= 
{m_S \over 2}{\rm Tr}\, S^2+
{\lambda_S \over 3} {\rm Tr}\, S^3+
{m_A \over 2} {\rm Tr} \, A^2+
\lambda {\rm Tr}\, A^2S + 
(m_\psi+\eta_A A)\psi\overline\psi\;.
\end{equation}

In complete analogy with case ({\it i}), the SO(10) symmetry forbids 
self-couplings for $A$ and $\psi$. They will be generated as higher 
dimensional terms, i.e. when $S$ is integrated out we get 
$A^4/\langle S\rangle$, and when $A$ is integrated out one gets 
$(\psi\overline\psi)^2/\langle A\rangle$. The pattern of symmetry 
breaking is taken again to be (\ref{sixteen}) with $\psi$ in the 
role of $\Sigma$, and the mass spectrum is obtained following a 
similar reasoning. However, the LR breaking is now performed by 
SU(2)$_R$ {\em doublets}, and the only state in 
$\psi, \overline\psi$ not participating in the super-Higgs 
mechanism is a neutral combination. The complete spectrum is now

\begin{center}
\framebox{\begin{tabular}{l|c}
\hspace{2cm}State & \hspace{1cm} Mass \\
\hline 
\begin{tabular}{l}
all of $S$ in ${\bf 54}$ \\
all of $A$ in ${\bf 45}$, except $(1,1,15)_A$
\end{tabular}
&  $\sim M_X$ \\
\hline
\begin{tabular}{l}
$(2,1,4)_\psi$  + $(2,1,\overline{4})_{\overline\psi}$ \\
color triplets of $(1,2,\overline{4})_\psi$  and 
$(1,2,{4})_{\overline\psi}$  \\
color triplets of $(1,1,15)_A$ 
\end{tabular}
&  $\sim M_C$ \\
\hline
\begin{tabular}{l}
one neutral and two charged combinations from \\
the color singlets of $(1,2,\overline{4})_\psi$ 
and $(1,2,{4})_{\overline\psi}$ 
\end{tabular}
& $\sim M_{R}$ \\
\hline
color octet and singlet of $(1,1,15)_A$ 
&  $\sim Max\left[ {M_R^2\over M_C}, {M_C^2\over M_X} 
\right]$ \\
\hline
\begin{tabular}{l}
neutral combination from the color singlets \\  
of $(1,2,\overline{4})_\psi$ and $(1,2,{4})_{\overline\psi}$ 
\end{tabular}
& $\sim {M_{R}^2 \over M_X}$
\end{tabular}}
\vspace{0.5cm}
\end{center}

Only the color octet and the extra Higgs doublets could play 
some role in the running. Our calculations show that the
only hope for unification is with
$M_C^2/M_X$ above $M_R$, but even in that case
the only extra 
non-singlet fields below $M_R$, the  heavy Higgs doublets, are 
not enough to change the usual MSSM picture, and no intermediate 
scales are present. 

\section{Conclusions}

The minimal supersymmetric standard model, if extrapolated 
to very high energies, predicts successfully the unification 
of gauge couplings. This would imply a desert above $M_W$ all 
the way to the unification scale at $10^{16}$ GeV, a rather 
dreadful scenario. Recently, there has been great excitement 
about the possibility of large compactified dimensions which 
would offer  plenty of new physics to fill in the desert, even
at scales as low as $1-10$ TeV \cite{extradim}. 
However, it is fair to say that the unification in this 
scenario \cite{ddg98} is still far from reaching the level of the 
usual supersymmetric field theory \cite{mssm}, and furthermore it is 
plagued by the severe proton decay problem.

Meanwhile, one is tempted to look for a possible 
oasis in the desert of energies between $M_W$ and $M_X$.
 We have argued here that the 
supersymmetric SO(10) theory offers naturally such an 
oasis. It allows for a new scale associated with 
a see-saw mechanism below $10^{14}$ GeV. Admittedly, 
this scale is too large to be of direct experimental 
interest, but it could play an important role for 
neutrino physics. Furthermore, it has a strong 
effect on proton decay suggesting a possibly dominant 
mode: $p \rightarrow \pi^0 e^+$. At the same time, 
in this theory R-parity is exact which solves the 
d=4 proton decay problem of low energy supersymmetry 
and guarantees the stability of the LSP (the lightest 
supersymmetric partner) \cite{amrs99}. This and other 
aspects of the theory will be discussed in a  
forthcoming publication. 

The main reason behind the existence of a new scale below 
$M_X$ is the violation of the survival principle as 
we have discussed at length. In many cases supersymmetric 
theories are characterized by the absence of cubic 
couplings in the superpotential; this leads to an 
important possibility of light states with masses of 
the order $M_I^2/M$, where $M_I$ 
is the scale of relevant symmetry and $M$ is the cut-off 
scale, as explained in the main body of this work. 
Such states, which often carry exotic quantum numbers,
thus may lie at a scale much lower than $M_I$ and possibly 
as low as the scales soon to be probed in new experiments.
Even if the reader is not excited by the examples we have provided, 
we hope that she or he will find even more interesting 
theories which will not suffer from the monotony of the 
desert and which will offer new physics much closer 
to the electro-weak scale. 

\acknowledgments
We thank Umberto Cotti, Gia Dvali and Hossein Sarmadi 
for useful discussions. 
The work of B.B. is supported by the Ministry of Science and 
Technology of the Republic of Slovenia; the work 
of A.R. and G.S. is partially supported  by EEC 
under the TMR contract ERBFMRX-CT960090 and that of A.M. by 
CDCHT-ULA Project No. C-898-98-05-B. A.R. is also supported by 
DOE grant No. DE-FG02-97ER41036. B.B. and A.M. thank ICTP for 
hospitality.

\end{document}